\documentclass{JHEP3}

  \usepackage{latexsym,bm,amsmath,amssymb,amsfonts}
  \usepackage{epsfig,graphics,graphicx,mathrsfs}
  \usepackage{slashed}
  \usepackage[latin1]{inputenc}

  \long\def\comment#1{ }

  \newcommand{\beq}{\begin{eqnarray}}
  \newcommand{\eeq}{\end{eqnarray}}
  
 \def\simge{\mathrel{%
   \rlap{\raise 0.511ex \hbox{$>$}}{\lower 0.511ex \hbox{$\sim$}}}}
\def\simle{\mathrel{
   \rlap{\raise 0.511ex \hbox{$<$}}{\lower 0.511ex \hbox{$\sim$}}}}

\vspace{1.5cm}

\title{\rm \LARGE
Twist analysis of the nucleon spin in QCD}

\author{Yoshitaka Hatta and Shinsuke Yoshida\\Faculty  of Pure and Applied Sciences, University
of Tsukuba, \\Tsukuba, Ibaraki 305-8571, Japan\\
E-mail: \email
{hatta@het.ph.tsukuba.ac.jp}, \email{yoshida@het.ph.tsukuba.ac.jp} }

\abstract{
The decomposition of the nucleon spin into that of quarks and gluons is related to twist--two GPDs according to Ji sum rule. Further decomposition into the helicity and the orbital angular momentum inevitably requires twist--three GPDs. In this paper we derive exact relations between twist--three GPDs and the canonical orbital angular momentum density of quarks and gluons, and check their consistency with the longitudinal spin sum rule. Our work demonstrates that the complete decomposition of the nucleon spin  fits well with the framework of perturbative QCD.   }

\begin{document}

\section{Introduction}

  In a by now classic paper \cite{Ji:1996ek}, Ji derived sum rules which relate certain moments of
generalized parton distributions (GPD) to quarks' and gluons' individual contributions to the nucleon spin $\frac{1}{2}\,$:
\beq
J^q = \frac{1}{2}\int_{-1}^1 dx\, x \left(H_q(x) +E_q(x)\right)\,, \qquad
J^g = \frac{1}{4} \int_{-1}^1 dx \left(H_g(x)+E_g(x)\right)\,. \label{11}
\eeq
Here, $J^q$ ($J^g$) is the \emph{total} spin of quarks (gluons), namely, it is the sum of the helicity and the orbital angular momentum (OAM). The corresponding GPDs, $H_{q,g}$ and $E_{q,g}\,$, are twist--\emph{two}, and hence represent the leading contributions to cross sections in  exclusive processes.\footnote{Our normalization of the gluon GPD is such that, in the forward limit, $H_g(x)=xG(x)$  where $G(x)$ is the gluon distribution function. } In the case of $J^q$, one can also extract the OAM of quarks $L^q$ in a longitudinally polarized nucleon by subtracting from $J^q$ the quark helicity $\frac{1}{2}\Delta \Sigma$ which can be determined from independent measurements. Thus, in Ji's framework the nucleon spin decomposition reads
\beq
\frac{1}{2} = J^q+ J^g =\frac{1}{2} \Delta \Sigma + L^q + J^g\,.  \label{jis}
\eeq
 Theoretical cleanness and experimental feasibility of this decomposition scheme has boosted the study of GPDs to the forefront  of present--day research on nucleon structure.

However, (\ref{11}) does not exhaust the whole content of the longitudinal spin structure. It turns out that one can access more detailed, interesting information if one goes to twist--\emph{three}. An early indication of this was seen in a parton model analysis \cite{Penttinen:2000dg,Hagler:2003jw} in which a particular twist--three GPD was shown to be related to $L^q$. [This will be reproduced in (\ref{full}) below in full QCD.] More recently, it has been recognized that a complete decomposition of the nucleon spin may be achieved at the twist--three level \cite{Hatta:2011ku,Ji:2012sj}, and this simultaneously resolves the longstanding controversy over whether such a decomposition is possible at all in QCD. [See the recent intense discussions in \cite{Chen:2008ag,Chen:2009mr,Wakamatsu:2010qj,Wakamatsu:2010cb,
Hatta:2011zs,Leader:2011za,Wakamatsu:2012ve,Ji:2012gc,Burkardt:2012sd,cedric}.]

Let us explicate this last point which has actually constituted the main motivation for the present study.
 Recall the well--known Jaffe--Manohar decomposition \cite{Jaffe:1989jz}
 \beq
\frac{1}{2}= \frac{1}{2} \Delta \Sigma + L^q_{can} + \Delta G + L^g_{can}\,, \label{jm}
\eeq
where $\Delta G$ is the gluon helicity, and $L^q_{can}$ and $L_{can}^g$ are the \emph{canonical} OAMs of quarks and gluons, respectively. They feature the canonical momentum (ordinary derivative) $\vec{x} \times i\vec{\partial}$  as opposed to Ji's dynamical OAM $L^q$ featuring the covariant derivative $\vec{x}\times i\vec{D}$. Because of this, each term of the decomposition (\ref{jm}) is not gauge invariant (excepting $\Delta \Sigma$) once the interaction is turned on. Nevertheless, starting from (\ref{jm}) and interpreting the fields as given in a particular, but  arbitrary  gauge, one can make the \emph{gauge invariant extension} (GIE)  of it \cite{Ji:2012sj,Ji:2012gc}. The first such attempt, starting from the light--cone (LC) gauge expression, was made by Bashinsky and Jaffe \cite{Bashinsky:1998if} which was recently rediscovered in a different line of argument \cite{Hatta:2011zs}. Another GIE based on the Coulomb gauge was suggested by the work of Chen \emph{et al.} \cite{Chen:2008ag,Chen:2009mr} which also prescribed a general procedure to construct a GIE.
 Operators obtained in any such GIE are gauge invariant by construction, so there is actually a plethora of physically (and numerically) inequivalent gauge invariant decomposition schemes.\footnote{Contrary to the claims in \cite{Wakamatsu:2010cb,Leader:2011za,Wakamatsu:2012ve}. See \cite{cedric} for the latest discussion.}
Still, the GIE based on the LC gauge \cite{Bashinsky:1998if,Hatta:2011zs,Hatta:2011ku} has a  quite distinguished status because it is the only GIE relevant to high energy experiments, as evidenced by the fact that $\Delta G$ obtained in this particular GIE (but not anything else) coincides with the experimentally measured gluon polarization. Therefore, when restricted to the context of high energy QCD, practically we do have a complete, gauge invariant decomposition of the nucleon spin. By introducing the so--called potential OAM
\beq
L_{pot} \equiv L^q- L_{can}^q\,, \label{we}
\eeq
one can make the following connection between (\ref{jis}) and (a GIE of) (\ref{jm})
\beq
J^q =\frac{1}{2}\Delta \Sigma + L_{can}^q+L_{pot}\,, \label{basic1}
\eeq
\beq
J^g+L_{pot}= \Delta G + L_{can}^g\,. \label{basic2}
\eeq
  In the case of the LC--gauge GIE, all the entries in (\ref{basic1}) and (\ref{basic2}) can be explicitly written as the matrix element of manifestly gauge invariant operators \cite{Hatta:2011ku}.

The goal of this paper is to understand (\ref{basic1}) and (\ref{basic2}) at the density level, $L_{can}^q=\int dx\, L_{can}^q(x)$, etc., where $x$ is the momentum fraction of partons.
We shall derive exact expressions for $L_{can}^q(x)$ and $L_{can}^g(x)$ in terms of the twist--two and newly defined twist--three GPDs. Via the equation of motion, the latter can be written as the sum of the twist--two part (`Wandzura--Wilczek contribution') and the `genuine twist--three', quark--gluon and three--gluon correlators. These results fit well with the standard framework of perturbative QCD, and therefore serve as the starting point for further first--principle calculations. We finally show the $x$--moments of the results and confirm that the lowest moment  reproduces (\ref{basic1}) and (\ref{basic2}).

\section{Quark canonical OAM}

 We start by defining the nonforward proton matrix element of the quark--gluon (`F--type') twist--three operator\footnote{Our conventions are $\gamma_5=-i\gamma^0\gamma^1\gamma^2\gamma^3$, $\epsilon^{0123}=+1$, and  $g^{\mu\nu}=\mbox{diag}(1,-1,-1,-1)$. The light--cone coordinates are defined as $x^\pm = \frac{1}{\sqrt{2}}(x^0\pm x^3)=x_\mp$. The four--vector indices are denoted by Greek letters $\alpha,\mu,...$  and the transverse coordinates (momenta) are denoted by the indices $i,j=1,2$. We also introduce the two--dimensional antisymmetric tensor $\epsilon^{+-ij}\equiv -\epsilon^{ij}=-\epsilon_{ij}\,$, $\epsilon^{12}=+1$. }
\beq
&&\int \frac{d\lambda}{2\pi} \frac{d\mu}{2\pi} e^{i\frac{\lambda}{2}(x_1+x_2)+i\mu(x_2-x_1)}
\langle P'S'|\bar{\psi}(-\lambda n/2)\gamma^+W_{-\frac{\lambda}{2}\mu}gF^{+i}(\mu n)W_{\mu\frac{\lambda}{2}}\psi(\lambda n/2)|PS\rangle \nonumber \\
&& \quad  = \frac{1}{2} \bar{P}^+ \epsilon^{+i \rho\sigma}\bar{u}(P'S')\gamma_5\gamma_\rho u(PS) \Delta_\sigma \Phi_F(x_1,x_2,\xi,t)+\cdots\,,  \nonumber \\
&&  \qquad \approx \bar{P}^+ \epsilon^{+i \rho\sigma}\bar{S}_\rho \Delta_\sigma \Phi_F(x_1,x_2)+\cdots\,, \label{1}
\eeq
\beq
&&\int \frac{d\lambda}{2\pi} \frac{d\mu}{2\pi} e^{i\frac{\lambda}{2}(x_1+x_2)+ i\mu(x_2-x_1)}
\langle P'S'|\bar{\psi}(-\lambda n/2)\gamma_5\gamma^+W_{-\frac{\lambda}{2}\mu}g\tilde{F}^{+i}(\mu n)
W_{\mu\frac{\lambda}{2}}\psi(\lambda n/2)|PS\rangle \nonumber \\
&&  \quad =  - \frac{i}{2}\bar{P}^+ \epsilon^{+i\rho\sigma} \bar{u}(P'S')\gamma_5\gamma_\rho u(PS) \Delta_\sigma \tilde{\Phi}_F(x_1,x_2,\xi,t) + \cdots\,, \nonumber \\
&& \qquad \approx - i\bar{P}^+ \epsilon^{+i\rho\sigma} \bar{S}_\rho \Delta_\sigma \tilde{\Phi}_F(x_1,x_2) + \cdots\,,  \label{2}
\eeq
 where  $P^\mu \approx \delta^\mu_+ P^+$ is the proton momentum taken in the infinite momentum frame in the 3--direction, and $S^\mu$ is the longitudinally polarized spin vector normalized as $P^2=-S^2=m^2$, the nucleon mass squared. We have also introduced the average $\bar{P}^\mu= \frac{1}{2}(P^\mu+P'^\mu) = (\bar{P}^+,0,0,\bar{P}^-)$ and a lightlike vector $n^\mu=\delta^\mu_-/\bar{P}^+$.  The Wilson line $W$ makes the nonlocal operators gauge invariant.\footnote{We shall use the same notation $W$ both for the fundamental and the adjoint Wilson lines. The subscripts of $W$ indicating the initial and final points  will be often omitted when they are obvious from the context.}
 The momentum transfer is denoted as  $\Delta^\mu = P'^\mu - P^\mu$ from which we define the skewness parameter $\xi \equiv -\Delta^+/2\bar{P}^+$  and the Mandelstam variable $t\equiv \Delta^2$. We   assume that $\Delta^\mu$ is small, and in the above expansion we have kept only the linear term in $\Delta$ which contains both the structure $\sim \epsilon^{ij}\Delta_j$ and a factor of $S^+=S_-$.\footnote{We note that the spinor product in (\ref{1}) and (\ref{2}) can be written in a seemingly different way using an identity
  \beq
  i\epsilon^{+i\rho\sigma} \bar{u}(P'S') \gamma_5\gamma_\rho u(PS) \Delta_\sigma =2\bar{P}^+\bar{u}(P'S')\gamma^i u(PS)\,,  \label{iden}
  \eeq
 which follows from the Dirac equation $0=(\slashed{P}-m)u(PS)=\bar{u}(P'S')(\slashed{P}'-m)$. The linear dependence on $\Delta$ (and $S^+$) is not manifest in this alternative form.
  }
  The latter is apparent in the third line of (\ref{1}) and (\ref{2}) where we approximated $S^\mu\approx  S'^\mu$ in the spinors (admissible due to the explicit factor of $\Delta_\sigma$) and denoted this common vector as  $\bar{S}^\mu$. By the same token, we neglected the dependence on $\xi$, $t$ in $\Phi_F$ and $\tilde{\Phi}_F$ (and in all the twist--three distributions defined later).
   In the following, we further assume that $\Delta^+=0$, namely the proton is elastically scattered. This turns out to be a convenient choice because, when calculating  the OAM, one differentiates  nonforward matrix elements with respect to $\Delta^{i=1,2}$ (but not $\Delta^+$) and then takes the limit $\Delta^\mu\to 0$ (see, e.g., (\ref{ji}) below). Thus one can actually set $\Delta^+=0$ from the beginning. As we shall see, in the discussion of the  gluon OAM this choice is also very helpful when relating the matrix elements of different operators via the equation of motion.

  From $PT$--invariance, we find the following symmetry properties
  \beq
 \Phi_F(x_1,x_2)=-\Phi_F(x_2,x_1)\,,
  \qquad \tilde{\Phi}_F(x_1,x_2)=\tilde{\Phi}_F(x_2,x_1)\,. \label{inv}
   \eeq
   As shown in \cite{Hatta:2011ku}, the function $\Phi_F(x_1,x_2)$ is related to the potential angular momentum (\ref{we})
 \beq
 L_{pot} = \int dx_1dx_2\,  {\mathcal P}\frac{1}{x_1-x_2} \Phi_F(x_1,x_2) = \int dXdx \,{\mathcal P} \frac{1}{x} \, \Phi_F(X,x)\,, \label{pote}
 \eeq
  where ${\mathcal P}$ denotes the principal value, and in the second equality we switched to the notation $X=\frac{x_1+x_2}{2}$, $x=x_1-x_2$.

Next define the `D--type' twist--three distributions
\beq
&&\int \frac{d\lambda}{2\pi} \frac{d\mu}{2\pi} e^{i\frac{\lambda}{2}(x_1+x_2)+i\mu(x_2-x_1)}
\langle P'S'|\bar{\psi}(-\lambda n/2)\gamma^+W_{-\frac{\lambda}{2}\mu}\overleftrightarrow{D}^i(\mu n)W_{\mu\frac{\lambda }{2}}\psi(\lambda n/2)|PS\rangle \nonumber \\
&& \qquad \approx
 \epsilon^{+i \rho\sigma}\bar{S}_\rho \Delta_\sigma \Phi_D(x_1,x_2)+\cdots\,,
\eeq
\beq
&& \int \frac{d\lambda}{2\pi} \frac{d\mu}{2\pi} e^{i\frac{\lambda}{2}(x_1+x_2)+i\mu(x_2-x_1)}
\langle P'S'|\bar{\psi}(-\lambda n/2)\gamma_5\gamma^+W_{-\frac{\lambda}{2}\mu}\epsilon^{+ij -}\overleftrightarrow{D}_j(\mu n)W_{\mu\frac{\lambda}{2}}\psi(\lambda n/2)|PS\rangle \nonumber \\
&& \qquad
\approx i\epsilon^{+i \rho\sigma}\bar{S}_\rho \Delta_\sigma \tilde{\Phi}_D(x_1,x_2)+\cdots\,,
\eeq
where  $\overleftrightarrow{D}^\alpha \equiv \frac{1}{2}(D^\alpha -\overleftarrow{D}^\alpha)$, $D^\alpha = \partial^\alpha + igA^\alpha$, $\overleftarrow{D}^\alpha=\overleftarrow{\partial}^\alpha -igA^\alpha$. The structure of the right--hand--side has been fixed by the same criteria and approximations as in (\ref{1}) and (\ref{2}).
 This time we have
\beq
\Phi_D(x_1,x_2)=\Phi_D(x_2,x_1)\,,
\qquad \tilde{\Phi}_D(x_1,x_2)=-\tilde{\Phi}_D(x_2,x_1)\,.
\eeq
The integral of $\Phi_D$ gives Ji's quark OAM $L^q$ \cite{feng}
\beq
L^q = \int dx_1dx_2 \Phi_D(x_1,x_2)= \frac{1}{2S^+}\lim_{\Delta \to 0} \frac{\partial}{i\partial \Delta^i}  \epsilon^{ij}   \langle P'S'|\bar{\psi}(0)\gamma^+ i\overleftrightarrow{D}^j(0) \psi(0)|PS\rangle\,.  \label{ji}
\eeq

There are generic relations between the F--type and D--type distributions \cite{Eguchi:2006qz} which can be derived from the following
 identities
\beq
W_{-\frac{\lambda}{2}\mu}D^i(\mu n)W_{\mu\frac{\lambda}{2}} = \frac{i}{P^+}\int_{\frac{\lambda}{2}}^\mu dt\,  W_{-\frac{\lambda}{2} t}\, gF^{+i}(tn)W_{t \frac{\lambda}{2}} + W_{-\frac{\lambda}{2} \frac{\lambda}{2}} D^i (\lambda n/2)\,, \nonumber \\
W_{-\frac{\lambda }{2}\mu}\overleftarrow{D}^i(\mu n)W_{\mu\frac{\lambda}{2}} = -\frac{i}{P^+}\int_{-\frac{\lambda}{2}}^\mu dt\,  W_{-\frac{\lambda}{2} t}\, gF^{+i}(tn)W_{t \frac{\lambda}{2}} + \overleftarrow{D}^i(-\lambda n/2) W_{-\frac{\lambda}{2} \frac{\lambda}{2}}\,.
\label{koi}
\eeq
Using these identities,  we find
 \beq
  \Phi_D(x_1,x_2) ={\mathcal P} \frac{1}{x_1-x_2} \Phi_F(x_1,x_2) + \delta (x_1-x_2) L_{can}^q(x_1)\,, \label{can}
  \eeq
  \beq
  \tilde{\Phi}_D(x_1,x_2) = {\mathcal P}\frac{1}{x_1-x_2}\tilde{\Phi}_F(x_1,x_2)\,, \label{rel}
  \eeq
  where
   \beq
   L_{can}^q(x) &\equiv & \frac{1}{2S^+} \lim_{\Delta \to 0} \frac{\partial}{i\partial \Delta^i}  \epsilon^{ij} \int \frac{d\lambda}{2\pi} e^{i\lambda x} \Biggl\{ \langle P'S'|\bar{\psi}(-\lambda n/2)\gamma^+ \nonumber \\ && \qquad \qquad \times \frac{1}{2}\left(W_{-\frac{\lambda}{2}\frac{\lambda}{2}} iD^j(\lambda n/2) -i\overleftarrow{D}^j(-\lambda n/2)W_{-\frac{\lambda}{2}\frac{\lambda}{2}}\right) \psi(\lambda n/2)|PS\rangle \nonumber \\
   && - \frac{1}{2\bar{P}^+}\int d\mu \, \frac{1}{2}\left(\epsilon (\mu-\lambda/2)+\epsilon(\mu+\lambda/2)\right) \nonumber \\
   && \qquad \qquad \times \langle P'S'|\bar{\psi}(-\lambda n/2)\gamma^+W_{-\frac{\lambda}{2}\mu}  gF^{+j}(\mu n)W_{\mu\frac{\lambda}{2}}\psi(\lambda n/2)|PS\rangle
   \Biggr\}\,.  \label{after}
   \eeq
   ($\epsilon(x)=x/|x|$ is the sign function.)
In (\ref{rel}), the delta function term $\delta(x_1-x_2)$ is not present because $\tilde{\Phi}_D$ is antisymmetric in $x_1$ and $x_2$. As already implied by the notation, the integral of $L_{can}^q$ gives quarks' canonical OAM
  \beq
 L_{can}^q\equiv  \int dx L_{can}^q(x) = \frac{1}{2S^+}\lim_{\Delta \to 0} \frac{\partial}{i\partial \Delta^i}  \epsilon^{ij}   \langle P'S'|\bar{\psi}(0)\gamma^+ i\overleftrightarrow{D}^j_{pure}(0) \psi(0)|PS\rangle\,,
  \eeq
   where
 \beq
 D^\alpha_{pure}(\lambda n)&\equiv& D^\alpha +\frac{ig}{\bar{P}^+} \int d\mu \,{\mathcal K}(\mu-\lambda)W_{\lambda\mu}F^{+\alpha}( \mu n) \nonumber \\
 &\equiv& D^\alpha-igA^\alpha_{phys}(\lambda n)  \nonumber \\
 &\equiv&   \partial^\alpha +igA_{pure}^\alpha(\lambda n)\,,
\label{phys}
\eeq
 with ${\mathcal K}(\mu)$ being either $\frac{1}{2}\epsilon(\mu)$ or $\pm \theta(\pm \mu)$,\footnote{Note that, in Eq.~(\ref{after}), one can replace $\frac{1}{2}\epsilon(\mu+ \lambda/2)=\pm(\theta(\pm (\mu+ \lambda/2))-\frac{1}{2})$ with $\pm \theta(\pm (\mu+ \lambda/2) )$ (similarly for $\frac{1}{2}\epsilon(\mu-\lambda/2)$) because the $\mu$--integral for the `$\pm \frac{1}{2}$' term is unrestricted, and this corresponds to the point $x_1=x_2$ in (\ref{1}) where the function $\Phi_F$ vanishes. The different choices of ${\mathcal K}$ correspond to different boundary conditions in the light--cone gauge. See \cite{Hatta:2011zs} for the detail.} is the gauge covariant generalization of the canonical momentum $i\partial^\alpha$ in the LC--gauge GIE \cite{Hatta:2011zs}. (\ref{phys}) also defines the `physical' and `pure gauge' parts of the gauge field $A^\alpha=A_{phys}^\alpha+A^\alpha_{pure}$ which play a key role in the construction of   \cite{Chen:2008ag,Chen:2009mr}.\footnote{The authors of \cite{Chen:2008ag,Chen:2009mr} work in the GIE based on the Coulomb gauge, and accordingly their $A_{phys}$ is different from ours. }
    We can thus regard  (\ref{can}) as the doubly--unintegrated version of the relation $L^q=L^q_{can} + L_{pot}$ corresponding to the decomposition $iD^\alpha=iD_{pure}^\alpha -gA_{phys}^\alpha$.    We see that at the density level, the decomposition of $L^q$ into the canonical and potential parts is very natural.

    Note also that, due to the delta function in (\ref{can}), the canonical OAM density has support only at $x_1=x_2$, namely, the gluon has zero energy and the outgoing and returning quarks have the same light--cone momentum in the quark--gluon  correlator. This implies that the argument $x$ of $L_{can}^q(x)$ can be indeed   interpreted as the momentum fraction of quarks, in much the same way as in the usual parton distribution function (PDF), and also suggests the uniqueness (or the preferred choice) in defining the density $L_{can}^q\equiv \int dx L_{can}^q(x)$. On the other hand, from this point of view, there seems to be an ambiguity in defining the density of the dynamical OAM (\ref{ji}), $L^q =\int dx L^q(x)$.\footnote{ For instance one may define $L^q(x_2) \equiv \int dx_1 \Phi_D(x_1,x_2)$, or $L'^q(X)\equiv \int d(x_1-x_2)\, \Phi_D(x_1,x_2)$ giving different functions $L^q(x) \neq L'^q(x)$ in general.} [See, also, the discussion in \cite{feng}.]

Let us derive a relation between the OAM and the generalized parton distributions (GPD).
From the equation of motion, one can show that
\beq
&&\bar{\psi}(-z^-/2)\gamma^i (WD_- - \overleftarrow{D}_-W)\psi(z^-/2) =
\bar{\psi}\gamma^+ (WD^i - \overleftarrow{D}^i W)\psi \nonumber \\
&& \qquad \qquad + i\epsilon^{ij}\bar{\psi}\gamma_5\gamma_j
(WD_- + \overleftarrow{D}_-W)\psi -i\epsilon^{ij}\bar{\psi}\gamma_5\gamma^+(WD_j + \overleftarrow{D}_jW)\psi\,.   \label{quark}
\eeq
(In  our convention, $D_- = D^+$.)
The nonforward matrix element of (\ref{quark}) reads
 \beq
&&  2\frac{\partial}{\partial z^-}\langle P'S'| \bar{\psi}(-z/2)\gamma^i W \psi(z/2) |PS\rangle =
\langle P'S'| \bar{\psi}\gamma^+ (WD^i - \overleftarrow{D}^i W)\psi |PS\rangle \nonumber \\
&& \qquad  + 2i\epsilon^{ij} {\mathcal D}_- \langle P'S'| \bar{\psi}\gamma_5\gamma_j
W\psi |PS\rangle -i\epsilon^{ij}\langle P'S'|\bar{\psi}\gamma_5\gamma^+(WD_j + \overleftarrow{D}_jW)\psi |PS\rangle\,.
\eeq
 The second term on the right--hand--side contains the total derivative ${\mathcal D}$ corresponding to the translation of spatial coordinates: ${\mathcal D}_\mu {\mathcal O}(-z,z)\equiv \lim_{a^\mu \to 0} \frac{1}{a^\mu}({\mathcal O}(-z+a,z+a)-{\mathcal O}(-z,z))$.
 This term vanishes since, in the nonforward matrix element, ${\mathcal D}^\mu \sim i(P'^\mu -P^\mu) =i\Delta^\mu$ and we assume $\Delta^+=0$. The matrix element of the left-hand-side may be parameterized by the twist--two \cite{Ji:1996ek} and twist--three \cite{Penttinen:2000dg} GPDs
\beq
&& \langle P'S'|\bar{\psi}(-z/2)\gamma^\mu W\psi(z/2)|PS\rangle = \int dx \, e^{-ix\bar{P}^+z^-}
\biggl\{ (H_q(x,\xi,t)+E_q(x,\xi,t))\bar{u}(P'S') \gamma^\mu u(PS) \nonumber \\
 &&  \qquad \qquad \qquad \qquad  - E_q(x,\xi,t)\frac{\bar{P}^\mu}{m}\bar{u}(P'S')u(PS) + G_3(x,\xi,t) \bar{u}(P'S')\gamma_\perp^\mu u(PS)+\cdots \biggr\}\,, \label{strik}
 \eeq
 where $\gamma_\perp^\mu = \delta^\mu_i \gamma^i$.    In (\ref{strik}), we have kept only the terms relevant to us.\footnote{There are different conventions for the parametrization of twist--three GPDs in the literature (cf., Refs.~\cite{Penttinen:2000dg,Kiptily:2002nx,Hagler:2003jw}). Here we take any one of such parametrizations, extract the structure $\bar{u}'\gamma_\perp^\mu u$ (by using (\ref{iden}), if necessary \cite{Hagler:2003jw}), and redefine its coefficient to be $G_3(x)$.   }  For the same reason as explained below (\ref{2}), the arguments $\xi$ and $t$ of GPDs will be neglected in the following.   Noting that (cf. (\ref{iden}))
 \beq
 \bar{u}(P'S') \gamma^i u(PS) \approx  i\epsilon^{ij}\Delta_j \frac{\bar{S}^+}{\bar{P}^+}\,, \label{lin}
 \eeq
and comparing the terms proportional to $\epsilon^{ij}\Delta_j \bar{S}^+$, we obtain
\beq
x(H_q(x)+E_q(x)+G_3(x) ) &=& \tilde{H}_q(x)+ \frac{1}{2}\int dx' \left(\Phi_D(x,x') + \Phi_D(x',x)
+\tilde{\Phi}_D(x,x') -\tilde{\Phi}_D(x',x) \right)\,, \nonumber \\
&=& \tilde{H}_q(x) +L_{can}^q(x) +\int dx' {\mathcal P}\frac{1}{x-x'} \left(\Phi_F(x,x') +\tilde{\Phi}_F(x,x')\right)\,,  \label{obtain}
\eeq
 where $\tilde{H}_q$ is another GPD which gives the polarized quark distribution $\tilde{H}_q(x,\Delta=0)=\Delta q(x)$ in the forward limit.
Integrating over $x$, we find, using (\ref{inv}) and (\ref{pote}),
\beq
\Delta q + L_{can}^q+L_{pot} = \int  x(H_q(x)+E_q(x)+G_3(x))dx\,.
\eeq
We now recall that $J^q=\frac{1}{2}\int x(H_q(x)+E_q(x)) dx = L^q+\frac{1}{2}\Delta q$. Then it follows that
\beq
\int dx\, xG_3(x) = -L^q\,.  \label{full}
\eeq
 This formula was previously derived in \cite{Penttinen:2000dg}, albeit in an approximation (`parton model') which neglects gluons.\\

To proceed further, we need another independent relation which can be used to eliminate $G_3$ from  (\ref{obtain}). For this purpose we employ the identity \cite{Balitsky:1987bk}
 \beq
&&   z^\mu\left(\frac{\partial}{\partial z^\mu} \bar{\psi}(-z/2)\gamma^\alpha W\psi(z/2) -\frac{\partial}{\partial z_\alpha} \bar{\psi}(-z/2)\gamma_\mu W\psi(z/2) \right)  \nonumber \\
&& \qquad = \frac{-i}{2} z^\mu \epsilon^{\alpha}_{\ \mu\nu\rho}{\mathcal D}^\nu \left(\bar{\psi}(-z/2)\gamma_5 \gamma^\rho W\psi(z/2) \right)
 \nonumber \\
 && \qquad \qquad \frac{-z^\mu}{2} \int_{-\frac{1}{2}}^{\frac{1}{2}} du \, \epsilon^{\alpha}_{\ \mu\nu\rho}\bar{\psi}(-z/2)\gamma_5\gamma^\rho WgF^{\nu\tau}(u z)z_\tau W\psi(z/2)  \nonumber \\
 &&  \qquad \qquad -i z^\mu \int_{-\frac{1}{2}}^{\frac{1}{2}} du\, u\, \bar{\psi}(-z/2) \gamma_\mu W gF^{\alpha\tau}(uz) z_\tau W\psi(z/2)\,.  \label{mast}
 \eeq
 In the above nonlocal operators, $z^\mu$ is generic,  not necessarily proportional to $n^\mu$.
The nonforward matrix element of the right--hand--side reads, in the case of $z^\mu=\delta^\mu_- z^-$ and for the component $\alpha=i$,
\beq
 && \frac{z^-}{2}\epsilon^{ij} \Delta_j \langle P'S'| \bar{\psi}(-z/2)\gamma_5 \gamma^+ W \psi(z/2) |PS\rangle  \nonumber \\
&&  \qquad +\frac{(z^-)^2}{2} \int_{-\frac{1}{2}}^{\frac{1}{2}} d\tau \, \langle P'S'|\bar{\psi}(-z/2)\gamma_5\gamma^+ W g\tilde{F}^{+i}(\tau z)W\psi(z/2)|PS\rangle  \nonumber \\
&&  \qquad +i (z^-)^2\int_{-\frac{1}{2}}^{\frac{1}{2}} d\tau\, \tau\, \langle P'S'| \bar{\psi}(-z/2)\gamma^+W gF^{+i}(\tau z) W \psi(z/2) |PS\rangle\,.
\eeq
The first term is related to the polarized quark distribution
\beq
  \frac{z^-}{2}\epsilon^{ij} \Delta_j \langle P'S'| \bar{\psi}(-z/2)\gamma_5 \gamma^+ W_{-\frac{1}{2},\frac{1}{2}} \psi(z/2) |PS\rangle
 = -i\epsilon^{ij}\Delta_j \frac{\bar{S}^+}{\bar{P}^+} \int dx\, e^{-ix\bar{P}^+z^-} \frac{d}{dx}\tilde{H}_q(x)\,, \nonumber \\ \eeq
 while the second and the third terms become
\beq
&&\frac{(z^-)^2}{2} \int_{-\frac{1}{2}}^{\frac{1}{2}} d\tau \, \langle P'S'|\bar{\psi}(-z/2)\gamma_5\gamma^+ W g\tilde{F}^{+i}(\tau z)W\psi(z/2)|PS\rangle
\nonumber \\
&&= -i\epsilon^{ij}\Delta_j \frac{\bar{S}^+}{\bar{P}^+} \int dx_1 e^{-ix_1\bar{P}^+z^-} \int dx_2\, {\mathcal P}\frac{1}{x_1-x_2} \left(\frac{\partial}{\partial x_1} +\frac{\partial}{\partial x_2}\right)\tilde{\Phi}_F(x_1,x_2)\,.
\eeq

\beq
&&i (z^-)^2\int_{-\frac{1}{2}}^{\frac{1}{2}} d\tau\, \tau\, \langle P'S'| \bar{\psi}(-z/2)\gamma^+W gF^{+i}(\tau z) W\psi(z/2) |PS\rangle
\nonumber \\
&&= -i\epsilon^{ij}\Delta_j \frac{\bar{S}^+}{\bar{P}^+} \int dx_1 e^{-ix_1\bar{P}^+z^-} \int dx_2\, {\mathcal P}\frac{1}{x_1-x_2} \left(\frac{\partial}{\partial x_1} -\frac{\partial}{\partial x_2}\right)\Phi_F(x_1,x_2)\,.
\eeq

Next consider the matrix element of the left--hand--side of (\ref{mast}). Away from the light--cone, (\ref{strik}) may be covariantly generalized to\footnote{Note that the GPD $E_q$  should be kept in the $\mu=+$ component because we are interested in the ${\mathcal O}(\Delta)$ terms.}
\beq
&&\langle P'S'|\bar{\psi}(-z/2)\gamma^\mu W\psi(z/2)|PS\rangle \nonumber \\
&& \qquad \qquad =\int dx\, e^{-ix\bar{P}\cdot z} \bar{u}(P'S')\left((H_q+E_q+G_3)\gamma^\mu -G_3 \frac{\gamma \cdot z}{\bar{P}\cdot z} \bar{P}^\mu \right)u(PS)+\cdots\,. \label{cov}
\eeq
This leads to
\beq
&&   z^\mu\left(\frac{\partial}{\partial z^\mu} \langle P'S'| \bar{\psi}(-z/2)\gamma^i W\psi(z/2)|PS\rangle -\frac{\partial}{\partial z_i} \langle P'S'|\bar{\psi}(-z/2)\gamma_\mu W\psi(z/2) |PS\rangle \right)
\nonumber \\
&& \qquad \approx -i \epsilon^{ij}\frac{\Delta_j \bar{S}^+}{\bar{P}^+} \int dx\, e^{-ix\bar{P}^+z^-}
 \left(\frac{d}{dx}(x(H_q(x)+E_q(x))) +x\frac{d}{dx}G_3(x)\right)\,,
\eeq
 where we have set $z^\mu = \delta^\mu_- z^-$ after differentiation and kept only the ${\mathcal O}(\Delta)$ terms using (\ref{lin}).
We thus find
\beq
 \frac{d}{dx}\bigl(x(H_q(x)+E_q(x))\bigr) +x\frac{d}{dx}G_3(x)&=& \frac{d}{dx}\tilde{H}_q(x)+ \int dx' \, {\mathcal P}\frac{1}{x-x'} \left(\frac{\partial}{\partial x} -\frac{\partial}{\partial x'}\right)\Phi_F(x,x')  \nonumber \\
 && +\int dx' \, {\mathcal P}\frac{1}{x-x'} \left(\frac{\partial}{\partial x} +\frac{\partial}{\partial x'}\right)\tilde{\Phi}_F(x,x')\,. \label{15}
\eeq
(\ref{15})  can be formally solved for  $G_3$ with the boundary condition $G_3(x=\pm 1)=0$. Substituting the result into (\ref{obtain}), and performing integration by parts, we arrive at
\beq
L_{can}^q(x)
&=& x\int_x^{\epsilon(x)} \frac{dx'}{x'} (H_q(x')+E_q(x')) -x\int_x^{\epsilon(x)} \frac{dx'}{x'^2} \tilde{H}_q(x') \nonumber  \\
&& \qquad  -
x\int_x^{\epsilon(x)} dx_1\int_{-1}^1 dx_2\Phi_F(x_1,x_2) {\mathcal P}\frac{3x_1-x_2}{x_1^2(x_1-x_2)^2}  \nonumber \\
&& \qquad -x\int_x^{\epsilon(x)} dx_1\int_{-1}^1 dx_2\tilde{\Phi}_F(x_1,x_2)  {\mathcal P}\frac{1}{x_1^2(x_1-x_2)}\,.
 \label{mom}
\eeq
This is an identity which relates the quark canonical OAM with the twist--two GPD and genuine twist--three quark--gluon correlators. By substituting (\ref{mom}) into (\ref{can}), one finds a similar identity for the density $\Phi_D(x_1,x_2)$ of Ji's quark OAM (cf., (\ref{ji}), see, also footnote 8). In the parton model (`Wandzura--Wilczek approximation'), one neglects the $\Phi$--terms \cite{Hagler:2003jw} after which there is no distinction between $L^q_{can}$ and $L^q$.

Finally in this section,  consider the moments of (\ref{mom}). Using the formula
\beq
\int_0^{\pm 1} dx\, x^{n-1} \int_x^{\pm 1} \frac{dz}{z}f(z) = \frac{1}{n}\int^{\pm 1}_0 dx\, x^{n-1} f(x)\,,
\eeq
 we find
 \beq
&& \int_{-1}^1 dx\, x^{n-1} L_{can}^q(x) = \frac{1}{n+1}\int_{-1}^1 dx\, x^n(H_q(x)+E_q(x))  - \frac{1}{n+1}  \int_{-1}^1 dx\,
 x^{n-1}\tilde{H}_q(x) \nonumber \\
 && \qquad \qquad - \frac{1}{n+1} \int_{-1}^1 dx dx'\, x^{n-1} \left({\mathcal P}\frac{3x-x'}{(x-x')^2}  \Phi_F(x,x')+{\mathcal P}\frac{1}{x-x'}\tilde{\Phi}_F(x,x')\right)\,.
 \eeq
 In particular, when $n=1$,
 \beq
 L_{can}^q= \int dx\,  L_{can}^q(x) &=&  \frac{1}{2}\int dx\, x(H_q(x)+E_q(x)) -\frac{1}{2} \int dx \tilde{H}_q(x)  \nonumber \\
 && - \frac{1}{2} \int dx dx' \left({\mathcal P}\frac{3x-x'}{(x-x')^2}  \Phi_F(x,x')+{\mathcal P}\frac{1}{x-x'}\tilde{\Phi}_F(x,x')\right) \nonumber \\
 &=& J_q -\frac{1}{2}\Delta q -L_{pot}\,, \label{cor}
 \eeq
 where we used (\ref{pote}) and the symmetry properties (\ref{inv}).
(\ref{cor}) is the correct spin decomposition formula (\ref{basic1}) in the quark sector.

\section{Gluon canonical OAM}

The analysis of the gluon canonical OAM is entirely analogous to the quark case.
 We first need several definitions.
\begin{itemize}

\item Unpolarized/polarized gluon distribution functions:
\beq
&&\int \frac{d\lambda}{2\pi}  e^{i\lambda x}\langle PS|F^{+\alpha}(0)WF^{+\beta}(\lambda n)|PS\rangle   \nonumber \\
&& = -\frac{1}{2}xG(x) (P^+)^2 (g^{\alpha\beta}-P^\alpha n^\beta -P^\beta n^\alpha)  -\frac{i}{2}x\Delta G(x)  P^+\epsilon^{+-\alpha\beta}S^+  + \cdots\,.
\eeq

\item Gluon GPDs:
\beq
&&\int \frac{d\lambda}{2\pi} e^{i\lambda x}\langle P'S'|-\frac{1}{2}\left(F^{\alpha\tau}(-\lambda n/2)WF^{\beta}_{\ \ \tau}(\lambda n/2) +F^{\beta\tau}WF^{\alpha}_{\ \, \tau}\right)+\frac{g^{\alpha\beta}}{4}F^{\rho\tau}WF_{\rho\tau}  |PS\rangle \nonumber \\
&& \qquad=\frac{1}{2}H_g(x)\bar{u}(P'S')\bar{P}^{(\alpha} \gamma^{\beta)} u(PS)   +\frac{1}{2}E_g(x)\bar{u}(P'S')\frac{\bar{P}^{(\alpha}i\sigma^{\beta)\tau}\Delta_\tau}{2m} u(PS)
\nonumber \\ && \qquad \qquad \qquad \qquad +\frac{1}{2}F_g(x) \bar{u}(P'S')\bar{P}^{(\alpha}\gamma_\perp^{\beta)} u(PS)   + \cdots\,,   \label{gpd}
\eeq
 where $A^{(\alpha}B^{\beta)}\equiv \frac{1}{2}(A^\alpha B^\beta +A^\beta B^\alpha)$. In addition to the usual twist--two GPDs $H_g$ and $E_g\,$,  we have introduced a  twist--three GPD $F_g$ which satisfies  $\int dx F_g(x)=0$.\footnote{ After the $x$--integration, the operator on the left--hand--side of (\ref{gpd}) becomes a local operator. Its matrix element should not depend on non-covariant expressions such as  $n^\mu$ and $\gamma^\mu_\perp$.}  There are other twist--three terms, but those are not needed for the present purpose.

\item F--type three--gluon correlators:
\beq
&&\frac{1}{(\bar{P}^+)^2}\int \frac{d\lambda}{2\pi}\frac{d\mu}{2\pi} e^{i\frac{\lambda}{2}(x_1+x_2)+i\mu(x_2-x_1)}\langle P'S'|F^{+i}(-\lambda n/2)WgF^{+j}(\mu n) W F^{+k}(\lambda n/2)|PS\rangle \nonumber \\
&& \approx
 \Bigl(M(x_1,x_2)g^{ik}\epsilon^{+j \rho\sigma} +M(x_2,x_2-x_1)g^{ij}\epsilon^{+k \rho\sigma}-M(x_1,x_1-x_2)g^{jk}\epsilon^{+i \rho\sigma}\Bigr)\bar{S}_\rho \Delta_\sigma + \cdots \,. \nonumber \\  \label{fty}
\eeq
In order to fix the tensorial structure of the right--hand--side, we followed the method developed for the three--gluon correlator in the transversely polarized case \cite{Ji:1992eu,Beppu:2010qn}. [As in (\ref{1}), $\bar{S}_\rho$ should more precisely read $\frac{1}{2}\bar{u}(P'S')\gamma_5\gamma_\rho u(PS)$.] By $PT$ invariance, it follows that
\beq
M(x_1,x_2)=-M(x_2,x_1)\,.  \label{they}
\eeq
On the other hand, from permutation symmetry we find
\beq
M(x_1,x_2)=M(-x_1,-x_2)\,.
\eeq
Contraction of $i$ with $k$ leads to an expression similar to (\ref{1})
\beq
&&\int \frac{d\lambda}{2\pi}\frac{d\mu}{2\pi} e^{i\frac{\lambda}{2}(x_1+x_2)+i\mu(x_2-x_1)} \langle P'S'|F^{+\alpha}(-\lambda n/2)WgF^{+j}(\mu n) W F^+_{\ \ \alpha}(\lambda n/2)|PS\rangle \nonumber \\
&& \qquad \approx (\bar{P}^+)^2
\epsilon^{+j\rho\sigma}\bar{S}_\rho \Delta_\sigma M_F(x_1,x_2) + \cdots\,,
\eeq
 where we defined
 \beq
M_F(x_1,x_2)&\equiv &2M(x_1,x_2)+M(x_2,x_2-x_1)-M(x_1,x_1-x_2)\,,
\eeq
with the property
$M_F(x_1,x_2)=-M_F(x_2,x_1)$. We also note the expression in the coordinate ($z^-$) space
\beq
&& \langle P'S'|F^{+\alpha}(\zeta n)WgF^{+j}(\mu n) W F^+_{\ \ \alpha}(\lambda n)|PS\rangle
\nonumber \\
 && \qquad \approx  (\bar{P}^+)^2 \epsilon^{+j\rho\sigma}\bar{S}_\rho \Delta_\sigma  \int dx_1dx_2 \, e^{-ix_1\lambda -i(x_2-x_1)\mu + ix_2\zeta}
  M_F(x_1,x_2)+ \cdots\,.
\eeq

 \item D--type three--gluon correlators:
\beq
&& \int \frac{d\lambda}{2\pi}\frac{d\mu}{2\pi} e^{i\frac{\lambda}{2}(x_1+x_2)+i(x_2-x_1)\mu} \langle P'S'|F^{+\alpha}(-\lambda n/2)W\overleftrightarrow{D}^i(\mu n) W F^+_{\ \ \alpha}(\lambda n/2)|PS\rangle \nonumber \\
&& \qquad \approx \bar{P}^+
\epsilon^{+i\rho\sigma}\bar{S}_\rho \Delta_\sigma M_D(x_1,x_2)+\cdots \,,
\eeq
 with
  $M_D(x_1,x_2)=M_D(x_2,x_1)$.
One may define, as in (\ref{fty}),  more general D--type correlators in which the Lorentz index $\alpha$ is not contracted. However, we do not need them for the present purpose.

\end{itemize}

In order to make contact with the gluon OAM, we need  three--gluon correlators in which one of the $F$'s is replaced by $A_{phys}$. They can be easily obtained via the convolution (\ref{phys}).
 \beq
&& \langle P'S'|F^{+\alpha}(\zeta n)WgF^{+i}(\mu n) W A^{phys}_{\alpha}(\lambda n)|PS\rangle \nonumber \\
 &&\qquad \approx  \bar{P}^+ \epsilon^{+i\rho\sigma}
 \bar{S}_\rho\Delta_\sigma \int dx_1dx_2 \, e^{-ix_1\lambda -i(x_2-x_1)\mu + ix_2\zeta}  i{\mathcal K}(x_1) M_F(x_1,x_2) +\cdots \,,
\eeq
\beq
&& \langle P'S'|F^{+\alpha}(\zeta n)W\overleftrightarrow{D}^i(\mu n) W A^{phys}_{\alpha}(\lambda n)|PS\rangle
\nonumber \\
&& \qquad \approx  \epsilon^{+i\rho\sigma}
\bar{S}_\rho\Delta_\sigma \int dx_1dx_2 \, e^{-ix_1\lambda -i(x_2-x_1)\mu+ ix_2\zeta }
i {\mathcal K}(x_1)  M_D(x_1,x_2)+\cdots\,,
\eeq
 where
 ${\mathcal K}(x)={\mathcal P}\frac{1}{x}$ in the case ${\mathcal K}(\mu)=\frac{1}{2}\epsilon(\mu)$ and
 ${\mathcal K}(x)=\frac{1}{x\mp i\epsilon}$ in the cases ${\mathcal K}(\mu)=\pm \theta(\pm \mu)$.
As in (\ref{can}) and (\ref{rel}),   the F--type and D--type distributions are related:
\beq
\frac{M_D(x_1,x_2)}{x_1} = \frac{M_F(x_1,x_2)}{x_1(x_1-x_2)} -\delta(x_1-x_2)L_{can}^g(x_1)\,. \label{using}
\eeq
The function $L_{can}^g(x)$ thus defined can be identified with the density of the gluon canonical OAM $\int dx\, L_{can}^g(x) = L_{can}^g$ because it satisfies \cite{Hatta:2011ku}\footnote{There is an overall sign error in the definition of the gluon canonical OAM in \cite{Hatta:2011ku}. The one--sided derivative $\overrightarrow{D}^\beta_{pure}$ in the definition of $L_g^{can}$ can be replaced by the symmetrized one  $\overleftrightarrow{D}^\beta_{pure}$ without affecting the value of $L_{can}^g$. }
\beq
i\epsilon^{+i\rho\sigma}\bar{S}_\rho \Delta_{\sigma} \int_{-1}^1 dx \, L_{can}^g(x)= -\langle P'S'|F^{+\alpha}\overleftrightarrow{D}^i_{pure} A_\alpha^{phys}|PS\rangle\,.
\eeq
Moreover, as already argued in the quark case,  the delta function $\delta(x_1-x_2)$ suggests the uniqueness (or the preferred choice) of the density $L_{can}^g \equiv \int dx L_{can}^g(x)$ with the variable $x$ interpretable as the momentum fraction of gluons.
\\

 Let us find the counterpart of (\ref{quark}) in the gluon case. Consider the following nonforward matrix element
\beq
I\equiv \frac{\partial}{\partial z^-} \langle P'S'|F^{+\tau}(-z^-/2)WF^{i}_{\ \ \tau}(z^-/2)+F^{i\tau}(-z^-/2)WF^+_{\ \ \tau}(z^-/2) |PS\rangle\,. \label{star}
\eeq
 Since $\Delta^+=0$ by assumption, we can freely make translation of the field coordinates in the $z^-$ direction and obtain
\beq
I&=&\langle P'S'|F^{+\tau}(0)W\overrightarrow{D}_-F^i_{\ \ \tau}(z^-)-F^{i\tau}(-z^-)\overleftarrow{D}_- WF^{+}_{\ \ \tau}(0)|PS\rangle \nonumber \\ &=& \langle P'S'|F^{+\tau}(0)W\left(\overrightarrow{D}^iF^+_{\ \ \tau}(z^-) -\overrightarrow{D}_\tau F^{+i}(z^-)\right) \nonumber \\ && \qquad \qquad  +\left(-F^+_{\ \ \tau}(-z^-)\overleftarrow{D}^i +F^{+i}(-z^-)\overleftarrow{D}_\tau\right)WF^{+\tau}(0)|PS\rangle\,,
\eeq
 where in the second equality we used the Jacobi identity.
We then use the formula
\beq
-F^{+\tau}(0)W\overrightarrow{D}_\tau F^{+i}(z^-)= F^{+\tau}\overleftarrow{D}_{\tau}WF^{+i}-\mathcal{D}_j\left(F^{+j}WF^{+i}\right)
\nonumber \\
-i\int_{0}^{z^-}d\omega^-\, F^{+j}(0) W gF^{+}_{\ \ j}(\omega^-)W F^{+i}(z^-)\,, \nonumber
\eeq
and  the equation of motion $D_\tau F^{\tau +}_a = g\bar{\psi} \gamma^+ t^a  \psi$
to get
\beq
I&=&\langle P'S'|\Biggl[ F^{+\tau}\left(W\overrightarrow{D}^i- \overleftarrow{D}^i W\right)F^+_{\ \ \tau} +\mathcal{D}_j \left(-F^{+j}WF^{+i} + F^{+i}WF^{+j}\right) \\
&& \qquad \qquad -\bar{\psi}(-z^-)\gamma^+W gF^{+i}(0)W\psi(-z^-) +\bar{\psi}(z^-)\gamma^+W gF^{+i}(0)W\psi(z^-) \nonumber \\
&&   +i\int_{-z^-/2}^{z^-/2} d\omega^- \left(-F^{+j}(-z^-/2)W gF^+_{\ \ j}(\omega^-)WF^{+i}(z^-/2) +F^{+i}W gF^+_{\ \ j}WF^{+j}\right) \Biggr]|PS\rangle\,. \nonumber
\eeq
These matrix elements can be expressed by the three--gluon correlators and $\Delta G(x)$. On the other hand, the starting expression (\ref{star}) can be directly expressed by the GPDs.
By equating the coefficient of $\bar{S}^+\epsilon^{ij}\Delta_j$ in these two expressions and using (\ref{using}), we find
\beq
&& \frac{1}{2}\bigl(H_g(x)+E_g(x)+F_g(x) \bigr) - \Delta G(x) +2\int dX\frac{\Phi_F(X,x)}{x} -2L_{can}^g(x)
\nonumber \\
&& \qquad \qquad =-2\int dx'\, {\mathcal P}\frac{M_F(x,x')}{x(x-x')} -2\int dx'\, {\mathcal P}\frac{\tilde{M}_F(x,x')}{x(x-x')}
\nonumber \\ && \qquad \qquad =
 -4\int dx'\, {\mathcal P}\frac{M(x,x')+M(x',x'-x)}{x(x-x')}\,, \label{sub}
\eeq
 where we defined a function
 \beq
 \tilde{M}_F(x,x') \equiv M(x,x-x')+M(x',x'-x)\,.
 \eeq
 Note that $\tilde{M}_F(x,x')$ is symmetric under $x\leftrightarrow x'$, although $M(x,x')$ and $M_F(x,x')$ are antisymmetric.
We now integrate (\ref{sub}) over $x$, recalling the basic properties
\beq
\frac{1}{2}\int_{-1}^1 dx (H_g(x)+E_g(x)) = 2J_g\,, \quad \int_{-1}^1 dx \Delta G(x) = 2\Delta G\,,
\quad  \int_{-1}^1 dx F_g(x)=0\,,
\eeq
as well as (\ref{pote}). The result is
\beq
J_g -\Delta G + L_{pot} - L_{can}^g &=& -2\int dxdx'\, {\mathcal P}\frac{M(x,x')+M(x',x'-x)}{x(x-x')}\,. \label{used}
\eeq
 Using the antisymmetric property (\ref{they}) of $M(x,x')$, one can easily check that the right--hand--side of (\ref{used}) is identically zero.
Eq.~(\ref{used}) is then nothing but the spin decomposition formula (\ref{basic2}).

 Next we consider the counterpart of (\ref{mast}) in the gluon case.
The generalization of (\ref{gpd}) off the light--cone $z^\mu \neq \delta^\mu_-z^-$ is  (cf.,(\ref{cov}))
\beq
&& z_\beta \langle P'S'|F^{\alpha \tau}WF^\beta_{\ \ \tau} +F^{\beta\tau}WF^\alpha_{\ \ \tau}  |PS\rangle =-\int dx\, e^{-ix\bar{P}\cdot z} \bar{u}(P'S') \nonumber \\
&& \qquad  \times \Bigl( (\bar{P}\cdot z \gamma^\alpha + \bar{P}^\alpha z\cdot \gamma)(H_g(x)+E_g(x)+F_g(x))-2\bar{P}^\alpha \gamma\cdot z F_g \Bigr) u(PS)+\cdots\,.
\eeq
This immediately gives
\beq
&& z^\mu \left(\frac{\partial}{\partial z^\mu} \,
  z_\beta\langle P'S'| F^{i \tau}WF^\beta_{\ \ \tau}+ F^{\beta\tau}WF^i_{\ \ \tau}|PS\rangle
-(i \leftrightarrow \mu)
\right) \nonumber \\
 &&\approx \frac{z^-}{2} \int dx\, e^{-ix\bar{P}^+z^-} \left\{\frac{d}{dx}\bigl(x(H_g(x)+E_g(x))\bigr)+x^2\frac{d}{dx}\frac{F_g(x)}{x}  \right\}i\epsilon^{i j}\Delta_j\bar{S}^+\,,
\label{lhs}
\eeq
 where again we set $z^\mu = \delta^\mu_{\, -} z^-$ after differentiation and kept only the linear terms in $\Delta$.
On the other hand, the left--hand--side of (\ref{lhs}) may be computed directly using the equation of motion and the Jacobi identity
\beq
&& z^\mu \left(\frac{\partial}{\partial z^\mu} \,
  z_\beta \langle P'S'| F^{i \tau}WF^\beta_{\ \ \tau}+ F^{\beta\tau}WF^i_{\ \ \tau} |PS\rangle
-(i \leftrightarrow \mu) \right)
\nonumber \\
&& =(z^-)^2 \langle P'S'| \biggl\{ -g\bar{\psi}(0)\gamma^+WF^{+i}(z)W\psi(0) +g\bar{\psi}(0)\gamma^+WF^{+i}(-z)W\psi(0)
\nonumber \\
&& \qquad \qquad \qquad -\mathcal{D}_j\left(F^{+j}WF^{+i}\right)+\mathcal{D}_j \left(F^{+i}WF^{+j}\right) \nonumber \\
 &&\qquad \qquad +iz^- \int_{-1/2}^{1/2}du \left(F^{+i}W gF^+_{\ \ j}(uz)WF^{+j}-F^{+j}W gF^+_{\ \ j}(uz)WF^{+i} \right)
\nonumber \\
&& \qquad \qquad \qquad \qquad \qquad +2i z^- \int_{-1/2}^{1/2} du\, u\, F^{+\tau}W gF^{+i}(uz)WF^{+}_{\ \ \tau} \biggr\}|PS\rangle \,. \label{25}
\eeq
[Again, $z^\mu = \delta^\mu_{\, -} z^-$ after differentiation.]
The matrix elements in (\ref{25}) can be evaluated as
\beq
&&z^-  \int dx \, e^{-ix\bar{P}^+z^-} \biggl\{-2\int dX \frac{\partial}{\partial x}\Phi_F(X,x) + \frac{d}{dx}(x\Delta G(x)) \nonumber \\&& \qquad \qquad  -2\int dx'\, {\mathcal P}\frac{1}{x-x'}\left(\frac{\partial}{\partial x}+\frac{\partial}{\partial x'}\right) \tilde{M}_F(x,x') \nonumber \\ && \qquad \qquad \qquad -2\int dx' \, {\mathcal P}\frac{1}{x-x'}\left(\frac{\partial}{\partial x}-\frac{\partial }{\partial x'} \right)M_F(x,x') \biggr\}i\epsilon^{ij}\Delta_j \bar{S}^+\,.
\eeq
Equating this with (\ref{lhs}), we obtain
\beq
 &&\frac{1}{2}\frac{d}{dx}\bigl(x(H_g(x)+E_g(x))\bigr)+\frac{x^2}{2}\frac{d}{dx}\frac{F_g(x)}{x}  =-2\int dX \frac{\partial}{\partial x}\Phi_F(X,x) + \frac{d}{dx}(x\Delta G(x)) \nonumber \\&&  \qquad \qquad \qquad \qquad \qquad \qquad \qquad \qquad -2\int dx' \, {\mathcal P} \frac{1}{x-x'}\left(\frac{\partial}{\partial x}+\frac{\partial}{\partial x'}\right) \tilde{M}_F(x,x') \nonumber \\ && \qquad \qquad \qquad \qquad \qquad \qquad \qquad  -2\int dx' \,{\mathcal P}\frac{1}{x-x'}\left(\frac{\partial}{\partial x}-\frac{\partial }{\partial x'} \right)M_F(x,x')\,,
\eeq
 which can be solved for $F_g(x)$. Eliminating $F_g$ from (\ref{sub}) in this way, we finally arrive at
\beq
L_{can}^g(x) &=& \frac{x}{2}\int_x^{\epsilon(x)} \frac{dx'}{x'^2}(H_g(x')+E_g(x'))- x\int_x^{\epsilon(x)} \frac{dx'}{x'^2} \Delta G(x') \nonumber \\
  && +2x\int_x^{\epsilon(x)} \frac{dx'}{x'^3} \int dX \Phi_F(X,x')  +2x\int_x^{\epsilon(x)}dx_1 \int_{-1}^1 dx_2 \tilde{M}_F(x_1,x_2) {\mathcal P}\frac{1}{x_1^3(x_1-x_2)}
\nonumber \\
 && \qquad +2 x\int_x^{\epsilon(x)} dx_1\int_{-1}^1 dx_2 M_F(x_1,x_2) {\mathcal P} \frac{2x_1-x_2}{x_1^3(x_1-x_2)^2}\,. \label{main}
 \eeq
This is our main result which expresses the gluon canonical OAM density in terms of twist--two GPDs and genuine twist--three contributions $\Phi_F$, $M_F$ and $\tilde{M}_F$. One may wish to neglect the latter in  the spirit of the Wandzura--Wilczek approximation. In that case it may make more sense to neglect only the last two terms $M_F$ and $\tilde{M}_F$ since they do not contribute to the integrated gluon OAM $L_{can}^g =\int L_{can}^g(x)$, see below.

In the moment space, (\ref{main}) takes the form
\beq
&& \int_{-1}^1 dx x^{n-1}L_{can}^g(x) = \frac{1}{2(n+1)}\int dx x^{n-1}(H_g(x)+E_g(x)) -  \frac{1}{n+1}\int dx x^{n-1}\Delta G(x) \nonumber \\ && \qquad \qquad \qquad \qquad \qquad  +\frac{2}{n+1}\int dx x^{n-2}\int dX \Phi_F(X,x)\nonumber \\
&&  \quad  +\frac{2}{n+1}\int dx  dx' x^{n-2}{\mathcal P}\frac{\tilde{M}_F(x,x')}{x-x'}  +\frac{2}{n+1}\int dx  dx' x^{n-2}{\mathcal P}\frac{2x-x'}{(x-x')^2} M_F(x,x')\,.
\eeq
In particular, when $n=1$, we get
\beq
L^g_{can}&=&J_g +L_{pot} -\Delta G +\int dxdx' \, {\mathcal P}\frac{\tilde{M}_F(x,x')}{x(x-x')} +\int dx dx'\, {\mathcal P}\frac{2x-x'}{x(x-x')^2}M_F(x,x')
\nonumber \\
&=& J_g +L_{pot}-\Delta G + \int dxdx' \, {\mathcal P}\frac{\tilde{M}_F(x,x')+M_F(x,x')}{x(x-x')} \,,
\eeq
where we used  the antisymmetric property of $M_F$. As already noted, the last integral vanishes and we recover the sum rule (\ref{basic2}).

\section{Conclusions}

Due to the difficulty in experimental measurements, twist--three GPDs have received relatively little attention so far, and even less attention in the particular context of longitudinal polarization. However, we have seen that they are intimately related to the decomposition of the nucleon spin beyond  (\ref{11}), and therefore provide strong motivation for further study. Especially it is an urgent task to pin down experimental processes which are sensitive to these distributions. In this respect, the connection \cite{Lorce:2011kd,Hatta:2011ku} between the canonical OAM and the OAM from the Wigner distribution (constructed from the `generalized parton correlation function' \cite{Meissner:2009ww})  may be helpful.

\section*{Acknowledgements}
We are very grateful to Feng Yuan for inspiring discussions and suggestions, and for informing us of his recent work with X.~Ji and X.~Xiong \cite{feng} which discusses related issues. The work of S.~Y. is supported by the Grant-in-Aid for Scientific
Research (No. 21340049).


\begin{thebibliography}{99}

\bibitem{Ji:1996ek}
  X.~D.~Ji,
  Phys.\ Rev.\ Lett.\  {\bf 78}, 610 (1997)
  [arXiv:hep-ph/9603249].


\bibitem{Penttinen:2000dg}
  M.~Penttinen, M.~V.~Polyakov, A.~G.~Shuvaev and M.~Strikman,
  Phys.\ Lett.\ B {\bf 491}, 96 (2000)
  [hep-ph/0006321].

\bibitem{Hagler:2003jw}
  P.~Hagler, A.~Mukherjee and A.~Schafer,
  Phys.\ Lett.\ B {\bf 582}, 55 (2004)
  [hep-ph/0310136].

\bibitem{Hatta:2011ku}
  Y.~Hatta,
  Phys.\ Lett.\ B {\bf 708}, 186 (2012)
  [arXiv:1111.3547 [hep-ph]].


\bibitem{Ji:2012sj}
  X.~Ji, X.~Xiong and F.~Yuan,
  arXiv:1202.2843 [hep-ph].





\bibitem{Chen:2008ag}
  X.~S.~Chen, X.~F.~Lu, W.~M.~Sun, F.~Wang and T.~Goldman,
  Phys.\ Rev.\ Lett.\  {\bf 100}, 232002 (2008)
  [arXiv:0806.3166 [hep-ph]].




\bibitem{Chen:2009mr}
  X.~S.~Chen, W.~M.~Sun, X.~F.~Lu, F.~Wang and T.~Goldman,
  Phys.\ Rev.\ Lett.\  {\bf 103}, 062001 (2009)
  [arXiv:0904.0321 [hep-ph]].

\bibitem{Wakamatsu:2010qj}
  M.~Wakamatsu,
  Phys.\ Rev.\  D {\bf 81}, 114010 (2010)
  [arXiv:1004.0268 [hep-ph]].

\bibitem{Wakamatsu:2010cb}
  M.~Wakamatsu,
  Phys.\ Rev.\ D {\bf 83}, 014012 (2011)
  [arXiv:1007.5355 [hep-ph]].





\bibitem{Hatta:2011zs}
  Y.~Hatta,
  Phys.\ Rev.\ D {\bf 84}, 041701(R) (2011)
  [arXiv:1101.5989 [hep-ph]].


\bibitem{Leader:2011za}
  E.~Leader,
  Phys.\ Rev.\  D {\bf 83}, 096012 (2011)
  [arXiv:1101.5956 [hep-ph]].


\bibitem{Wakamatsu:2012ve}
  M.~Wakamatsu,
  arXiv:1204.2860 [hep-ph].

\bibitem{Ji:2012gc}
  X.~Ji, Y.~Xu and Y.~Zhao,
  arXiv:1205.0156 [hep-ph].



\bibitem{Burkardt:2012sd}
  M.~Burkardt,
  arXiv:1205.2916 [hep-ph].



\bibitem{cedric}
  C.~Lorc\'e,
  arXiv:1205.6483 [hep-ph].


\bibitem{Jaffe:1989jz}
  R.~L.~Jaffe and A.~Manohar,
  Nucl.\ Phys.\  B {\bf 337}, 509 (1990).

\bibitem{Bashinsky:1998if}
  S.~Bashinsky and R.~L.~Jaffe,
  Nucl.\ Phys.\ B {\bf 536}, 303 (1998)
  [hep-ph/9804397].







\bibitem{feng} F.~Yuan, private communications;
  X.~Ji, X.~Xiong and F.~Yuan,
  arXiv:1207.5221 [hep-ph].


\bibitem{Eguchi:2006qz}
  H.~Eguchi, Y.~Koike and K.~Tanaka,
  Nucl.\ Phys.\ B {\bf 752}, 1 (2006)
  [hep-ph/0604003].

\bibitem{Kiptily:2002nx}
  D.~V.~Kiptily and M.~V.~Polyakov,
  Eur.\ Phys.\ J.\ C {\bf 37}, 105 (2004)
  [hep-ph/0212372].


\bibitem{Balitsky:1987bk}
  I.~I.~Balitsky and V.~M.~Braun,
  Nucl.\ Phys.\ B {\bf 311}, 541 (1989).







\bibitem{Ji:1992eu}
  X.~-D.~Ji,
  Phys.\ Lett.\ B {\bf 289}, 137 (1992).


\bibitem{Beppu:2010qn}
  H.~Beppu, Y.~Koike, K.~Tanaka and S.~Yoshida,
  Phys.\ Rev.\ D {\bf 82}, 054005 (2010)
  [arXiv:1007.2034 [hep-ph]].



\bibitem{Lorce:2011kd}
  C.~Lorc\'e and B.~Pasquini,
  Phys.\ Rev.\  D {\bf 84}, 014015 (2011)
  [arXiv:1106.0139 [hep-ph]].

\bibitem{Meissner:2009ww}
  S.~Meissner, A.~Metz and M.~Schlegel,
  JHEP {\bf 0908}, 056 (2009)
  [arXiv:0906.5323 [hep-ph]].


\end{thebibliography}
\end{document}